\documentclass[aip,rsi,floatfix,superscriptaddress,reprint]{revtex4-1}

\usepackage{graphicx}
\usepackage{amsmath}

\newcommand{\Rb}{$^{87}$Rb}
\newcommand{\pRb}{p_\text{Rb}} % partial pressure
\newcommand{\betaRb}{\beta_\text{Rb}} % proportionality const. for
\newcommand{\Nfull}{N_\text{f}} % atom nb in fully loaded MOT
\newcommand{\tauMot}{\tau_\mathit{MOT}}     % time used for MOT loading

\newcommand{\Nmot}{N}       % MOT atom number
\newcommand{\Tmot}{t_M}     % time used for MOT loading

\newcommand{\be}{\begin{equation}}
\newcommand{\ee}{\end{equation}}

\begin{document}

\title{Alkali vapor pressure modulation on the 100\,ms scale in a
  single-cell vacuum system for cold atom experiments}

% repeat the \author .. \affiliation  etc. as needed
% \email, \thanks, \homepage, \altaffiliation all apply to the current author.
% Explanatory text should go in the []'s, 
% actual e-mail address or url should go in the {}'s for \email and \homepage.
% Please use the appropriate macro for the type of information

% \affiliation command applies to all authors since the last \affiliation command. 
% The \affiliation command should follow the other information.

\author{Vincent Dugrain}
\affiliation{Laboratoire Kastler Brossel, ENS, UPMC, CNRS, 24 rue Lhomond, 75005 Paris, France}

\author{Peter Rosenbusch}
\affiliation{LNE-SYRTE, Observatoire de Paris, CNRS, UPMC, 61 av de l'Observatoire, 75014 Paris, France}
%\author{Christian Deutsch}
%\affiliation{Laboratoire Kastler Brossel, ENS, UPMC, CNRS, 24 rue Lhomond, 75005 Paris, France}

\author{Jakob Reichel}
\affiliation{Laboratoire Kastler Brossel, ENS, UPMC, CNRS, 24 rue Lhomond, 75005 Paris, France}

%\email[]{Your e-mail address}
%\homepage[]{Your web page}
%\thanks{}
%\altaffiliation{}

\date{\today}

\begin{abstract}
  We describe and characterize a device for alkali vapor pressure
  modulation on the 100\,ms timescale in a single-cell cold atom
  experiment. Its mechanism is based on optimized heat conduction
  between a current-modulated alkali dispenser and a heat sink at room
  temperature. We have studied both the short-term behavior during
  individual pulses and the long-term pressure evolution in the
  cell. The device combines fast trap loading and relatively long trap
  lifetime, enabling high repetition rates in a very simple setup.
  These features make it particularly suitable for portable atomic
  sensors.
\end{abstract}

% insert suggested PACS numbers in braces on next line

\maketitle %\maketitle must follow title, authors, abstract and \pacs

\section{Introduction}
% As cold atoms have become a tool for precision experiments and
% metrology \cite{}, there is a need for compact and simple cold atom
% sources with high repetition rates.
Magneto-optical trap (MOT) loading is a limiting factor to the
repetition rate of cold atom experiments. Fast loading is crucial for
practical applications of cold atoms, such as atom interferometry
\cite{Cronin09}, as well as for experiments with high data volume,
such as quantum state tomography of multiparticle entangled states
\cite{Riedel10,Haas14}. Ideally, the MOT loading time $\Tmot$ should
be a small contribution to the cycle time, which means that it should
be shorter than about 1\,s in experiments involving evaporative
cooling. The MOT loading rate $R$ is proportional to the partial vapor
pressure of the element to be cooled (\Rb\ in this
article). Consequently, this pressure should be high for fast loading,
but low for long lifetimes after loading.  A widespread approach is to
use a two-chamber vacuum system with differential pumping. However,
this considerably increases the complexity of the vacuum and optical
systems. An interesting alternative is to temporally modulate the
vapor pressure inside a simple, single-cell apparatus. Light-induced
atom desorption (LIAD) from the cell surfaces has been investigated
for this purpose
\cite{Anderson01,Haensel01a,Atutov03,Du04,Klempt06,Mimoun10}. Although
it performs well in some experiments, there can be reproducibility
issues, at least for Rb. The pressure modulation factor and also the
effective capacity of the surfaces (i.e., the number of pulses before
depletion) vary by orders of magnitude between experiments. Strong
dependence on material properties such as surface contamination is a
possible explanation \cite{Anderson12PC}. Another, straightforward way
to modulate the pressure of alkali vapors is to modulate the heating
current in commercial alkali dispensers \cite{Wieman95}. A fast
pressure rise is easily achieved by using elevated heating
currents. However, the timescale of pressure {\em decay}
%in published results is 
%above 1\,s \cite{McDowall12}, and 
is typically between 3 and $4\,$s
\cite{Wieman95,Fortagh98a,Rapol01,Fortagh03,Bartalini05,Scherer12}, too long to
achieve a significant reduction of the cycle time.
%, with the exception of \cite{Griffin05}, discussed below. 
Here we show that
this decay can be accelerated by more than an order of magnitude
without sacrificing the simplicity and safe operation of a dispenser
setup, by optimizing the thermal contact of the dispenser to an outside
heatsink.  Our device achieves pressure decay
constants on the 100\,ms timescale as desired, well within the range
of interest for high repetition rate experiments.

The long-term pressure evolution in a pulsed-source system is
influenced by slow processes such as adsorption-desorption dynamics,
which can introduce timescales as long as several hours. Therefore, we
also investigate the long-term evolution of \Rb\ pressure, background
pressure and number of trapped atoms. As expected, the background
pressure increases with the intensity of the dispenser pulses. We find
evidence for both adsorption-desorption dynamics and slow heat-up of
the vacuum apparatus. The latter effect is specific to our setup,
which was not designed for this purpose, and we discuss
straightforward improvements. In spite of these effects, we find that
background pressure in our system remains lower than the
pressure in a similar system with traditional, continuously heated
dispenser, while the pulsed source accelerates MOT loading by more
than an order of magnitude.
%The ultimate limits to the pressure modulation factor
%are likely to be set by the adsorption dynamics, which depends on the
%choice of cell materials and cell geometry. 
%the trapped-atom clock on a chip, TACC \cite{Deutsch10}, 
This makes our system very attractive for many applications of cold
alkali atoms.

% When the source turns off so fast, two other effects may dominate the
% pressure evolution: pumping speed and the adsorption-desorption
% dynamics of the vapor on the cell surfaces. In a small cell like ours
% with a volume $V$ of a few hundred cubic centimeters and a
% conduction-limited pumping speed $S$ of a few $\mbox{l}/\mbox{s}$, the
% nominal pumping time constant $\tau_p=V/S$ is in the 100\,ms range,
% which is acceptable. However, the material-dependent sticking time of the
% alkali and other atoms on the cell walls can introduce a slow
% evolution of the base pressure. We have performed long-term
% measurements which characterize our system and allow some conclusions
% about limiting factors.

The outline of this paper is as follows.  We first discuss thermal
coupling of the dispenser to the environment, motivate why an optimum
range exists for this coupling, and estimate that the coupling found
in usual experiments is too weak. Section~\ref{sec:adsorption}
collects relevant facts about the mechanisms and time constants that
govern the pressure evolution in the cell. In
section~\ref{sec:experiment}, we describe our setup, where the thermal
coupling is improved by copper parts which embrace the dispenser shell
and thermally connect it to an outside
heatsink. Sections~\ref{sec:shortTerm} and \ref{sec:longTerm} present
and discuss the experimental results obtained when this dispenser is
used as a pulsed source for a MOT. It is shown that a steady-state
regime can be established and its characteristics are
investigated. Finally, section~\ref{sec:loadingRate} investigates the
features of this steady-state regime as a function the alkali pulse
intensity.
% The dynamics of absorption of the atoms on the cell surfaces are
% shown to limit the maximum rubidium pulse intensity that can be used
% for cold atoms experiments.

%In the last section, we describe complementary measurements,
% including attempts to use a heating laser instead of resistive
% heating, and open-shell dispensers.

\section{Pressure modulation time constants and dispenser heat conduction}
\label{sec:timeConstants}
% Our dispenser holder, described in the next section, accelerates
% pressure decay by improving thermal conduction from the dispenser to
% a cold reservoir outside of the vacuum system. To motiv
% The experimental results in
% sec.~\ref{sec:} show that this indeed accelerates the pressure decay
% after the pulse to the 100\,ms timescale as desired.
Alkali metal release from dispensers is a thermally activated process
with a strongly nonlinear temperature
dependence. Because of the high temperature involved, dispensers also
release some ``dirt'', such as water vapor \cite{Gruener09}. Cold atom
experiments typically use commercial dispensers (SAES getters
``FT-type'' series) with constant heating currents between 3.5 and
4\,A, corresponding to dispenser temperatures of about $500^\circ$C
according to the manufacturer. To reach pressure modulation time
constants in the 100\,ms range, a first condition is that vapor
release from the dispenser be modulated on this timescale, or faster.
For our purposes, the dispenser must be brought to the desired
temperature $T_1\sim 500^\circ$C in less than 100\,ms, kept there for
the MOT loading time, and then must cool down to significantly below
$T_1$ in less than 100\,ms. Fast heating is easily accomplished by
applying a high-current pulse instead of continuous heating at lower
current.  The cooling timescale, however, is determined by the dispenser's
thermal coupling to the outside world.

We have measured $R=0.055\,\Omega$ for the resistance of the 12\,mm
long dispenser used in our experiments, so that the typical constant
currents correspond to heating powers $P=0.7\ldots 0.9\,$W.  In usual
experiments, radiation and heat conduction through the connecting
wires both contribute to the (inefficient) cooling of the
dispenser. The following estimate gives an idea of the orders of
magnitude. An upper limit of the dispenser's heat capacity $C_d$ can
be obtained from the easily measurable electrical heating power and
heating time to reach alkali emission when heating rapidly from room
temperature. Assuming that emission starts around $500^\circ \,$C, we
obtain $C_d\le 0.07\,\mbox{J}/\mbox{K}$ from our experimental
data. (Some heat is lost to the mount during heating, which is why
this method yields an upper limit only.)  The power lost by radiation,
$P_{\text{rad}}$, can be estimated from the dispenser
surface. However, the result depends on the emissivity $\epsilon$ of
the dispenser shell, which can be anywhere between $0.8$ and below
$0.1$ for shiny metal. With $\epsilon=0.8$, one obtains
$P_{\text{rad}}\approx 0.8\,W$ at $500^\circ$C for our dispenser
size. The thermal conductance $h$ of the connecting wires is
considerable -- $h=5\,\mbox{mW}/\mbox{K}$ for a 1\,mm diameter copper
wire of 20\,cm length -- but is throttled by the mechanical connection
between the dispenser and the wire. Given that $P<1\,$W, the effective
$h$ must be $h\lesssim 2\,\mbox{mW}/\mbox{K}$, because otherwise the
temperature of $500^\circ$C could not be reached.
% (Their radiation loss is negligible).
% This results in $\tau\approx 14\,$s. Being larger than the 3\,s
% timescale of pressure decay, this value is compatible with the
% experiments given the strong nonlinear pressure-temperature
% dependence. Unfortunately, we cannot make a more quantitative
% comparison at this stage.

To accelerate dispenser cooldown, the thermal conductance $h$ from the
dispenser to the room-temperature reservoir must be significantly
increased. The contribution from radiation loss can then be neglected,
and we can model the temperature evolution with a simple heat equation:
\begin{equation}
  C_d \dot{T} = P - h(T-T_0)\,.  
\end{equation}
Here, $T_0$ is the reservoir temperature and we have taken the
dispenser temperature to be homogeneous. For cooling ($P=0$) from
an initial temperature $T(0)=T_1$, this equation leads to $T(t) = T_0 +
(T_1-T_0)\exp(-t/\tau)$ with
\begin{equation}
\tau=\frac{C_d}{h}\,.
\end{equation}
Improving the heat conduction $h$ has the desired effect of reducing
$\tau$. However, we also need to consider the undesired heat flow from
the dispenser to the reservoir during the heating phase, which
increases with $h$.  In our simple model, for heating with constant
power and $T(0)=T_0$, the temperature rise is $T(t)=T_0+
\frac{P}{h}(1-\exp(-t/\tau))$. The maximum temperature that can be
obtained is 
\begin{equation}
  T(\infty)-T_0=\frac{P}{h}\,.
\end{equation}
Therefore, as a first condition, $P$ must be large enough to maintain
$P/h>T_1-T_0$. The heat flow to the reservoir during the heating phase
is $\Phi_l(t) = P (1-\exp(-t/\tau))$ and the total energy leaking to
the reservoir during heating to $T_1$ is $\frac{P
  C_d}{h}\ln(\frac{1}{1-(T_1-T_0)h/P})-(T_1-T_0)C_d$.  It becomes
neglegible for $P\gg h(T_1-T_0)=\frac{C_d}{\tau}$. (This can also be
seen by noting that $\Phi(t)$ remains small for $t\ll\tau$, and
requiring that the heating time $t_1$ be $t_1\ll\tau$.)  This shows
that reservoir heating during dispensing can be kept small even for
large $h$, by using a high heating power and a short heating
time. Nevertheless, it would be wrong to conclude that $h$ should be
made as large as possible. First, larger $h$ requires proportionally
increased $P$, and $P$ is limited in practice. Moreover, an important
practical limit comes from the fact that the required current exceeds
the steady-state destruction threshold: when such a current is applied
for too long, the dispenser melts like a fuse. The higher the current,
the shorter the safety margin to destruction. With our holder
described below, the required heating current is above 20\,A. We
have not made a systematic study of the destruction current, but we
know from an unintentional experiment that a current of $40\,$A
destroys the dispenser in less than 2\,s. (A continuous current of
20\,A is possible, though not desirable as it empties the dispenser in
less than two days -- another unintended experimental result.)

% To be more quantitative, we have made an approximate measurement of
% the heat capacity of one of our dispensers (SAES FT type, 12\,mm
% active length) by glueing a small thermistor to the dispenser shell
% and monitoring heating voltage, current and temperature rise. This
% gave an upper limit (because of heat capacity of the thermistor) of
% $C_d<90\,\mbox{mJ}/\mbox{K}$. Note that this value is largely
% dominated by the heat capacity of the shell, as can be
% estimated from its dimensions and the specific heat capacity of
% the shell material, which is very probably nichrome.
% % Delta T >= 400K => E>=36J => P>=72W to supply this E in 0.5s.

%Because of the nonlinear dependence of the dispenser's alkali emission
%on temperature, the model does not directly yield the time constant of
%pressure decay, but only an upper limit. 
We conclude that an optimum regime exists
for the thermal coupling $h$. It is situated between the usual,
undercoupled, slow-cooling regime and an overcoupled regime with
strong reservoir heating and/or high likelihood of dispenser
overheating. Our dispenser mount brings $h$ into the optimum regime.
Before proceeding, we note that the heat conduction model also
provides an explanation for the slightly improved pressure decay in
\cite{McDowall12} and for the very fast decay observed in the
laser-heated dispenser experiment \cite{Griffin05} for low heating
energies. In \cite{McDowall12}, the dispenser sleeves are attached to
copper rods instead of wires, and the resulting pressure decay after
dispenser turn-off has an initial decay constant of 1.8\,s, about a
factor of 2 faster than in experiments with wire-mounted dispensers
\cite{Fortagh98a,Rapol01,Bartalini05}.  In \cite{Griffin05}, the
dispenser shell was attached to a ceramic plate over its full length,
providing much higher $h$ than wire mounting. Additionally, only a
fraction of the dispenser surface was laser heated so that the rest of
the dispenser effectively acted as a well-coupled reservoir. The
authors place an upper limit of 100\,ms on the switch-off time. From
the schematic shown in the article, it seems that the ceramic plate
itself did not have a substantial thermal contact to the outside,
which would provide an explanation for the much slower decay they observed
for high heating powers. Another possibe explanation would be
adsorption-desorption dynamics, treated below.

\section{Time constants of pumping and adsorption-desorption dynamics}
\label{sec:adsorption}
Rapidly modulating the dispenser itself is not enough.  Additionally,
the timescale for evacuating excess alkali and dirt vapor after MOT
loading must also be fast. If one considers the
standard conductance formulas for the molecular flow regime, it seems
easy to achieve a sufficiently fast pumping time constant, even with
fairly small pumps. For our apparatus described below,
the calculated pumping time constant (including tube lengths and
angles) is 70\,ms.
%It would be easy to further reduce this time with
%larger-diameter tubing and/or an additional getter pump. 
However, these formulas describe an ideal gas and do
not take into account the adsorption and desorption processes that
occur for the highly reactive alkali vapor.

The binding energies and sticking times of alkali metals on UHV cell
surfaces vary widely depending on the surface material and on the
amount of coverage. It is difficult to obtain reliable data on the
time constants and the situation appears to be particularly obscure
for Rb, perhaps because its melting point is so close to room
temperature. Nevertheless, some experimental facts are well
established.  First, there is the phenomenon sometimes called
``curing'': when alkali vapor is first introduced from a reservoir
into a freshly baked cell, it takes a long time, often several days,
until a substantial vapor pressure can be created throughout the cell.
This is most clearly observed in long UHV tubes or elongated cells
\cite{Ma09}, but the phenomenon is general \cite{Wieman95} and we have
observed it in all our experiments.
% As an example, in the experiment described in
% \cite{Reichel99}, a Rb reservoir was connected through a valve and
% approximately 30\,cm of steel tubing with a glass cell. After opening
% the valve and heating the reservoir, it took several days until any Rb
% fluorescence could be observed in the glass cell.
Once a significant vapor pressure has been reached in the cell, it can
then be adjusted on a much shorter timescale (tens of minutes) by
adjusting the source rate, via a reservoir valve for example.  The
curing phenomenon suggests that clean surfaces form strong bounds with
the impinging alkali atoms, whereas binding is much weaker once the
surface has adsorbed a certain amount (often suggestively called a
``monolayer'').
% If the reservoir is removed and the tube is pumped, the vapor
% pressure decreases very slowly, over several days at room
% temperature and faster if the tube is heated \cite{Wieman95}.
%A quantitative investigation of the ``curing'' phenomenon using a long
%cylindrical Pyrex cell is described in \cite{Ma09}. The phenomenon is
%material-dependent. 
%According to \cite{Wieman95}, fused silica and
%Pyrex behave similarly to stainless steel, while other types of
%glass (microscope slides, for example), continue to ``pump'' Rb even
%after extensive exposure to Rb vapor.
The main consequence of the curing phenomenon in our context is that
data should only be taken after the transitory curing phase has
ended. All data presented below has been taken after several months of
dispenser operation and fulfills this condition.

% However, the cured surface still does not behave like bulk Rb.
% Otherwise, the pressure in a cured cell would be as high as if the
% cell contained a macroscopic drop of Rb metal (i.e., reaching the
% saturated vapor pressure if the cell is not pumped), whereas in
% reality it is much lower.
The cured surface still has a memory effect for its recent alkali
exposure.
% Thus, for example, when a
% dispenser source is shut off, 
% a nearby pressure gauge may indicate a
% fast initial pressure drop, but then a much slower decrease over
% several hours. 
It is well known, for example, that
%the atom number in a MOT needs a long time (tens of minutes or more)
%to stabilize to a new value when the dispenser current is
%changed. Also, 
a small MOT can still be loaded for many hours after the source has
been turned off. These timescales are too long to be explained by
dispenser cool-down and suggest that some fraction of the alkali atoms
stick to the cell walls and thus spend an average time $\tau_s$ (on the
order of hours) inside the cell before being pumped away. To estimate
the consequences of this effect, consider the following very crude
model.

Two effects contribute to the alkali pressure in the cell
$p(t)$. The dispenser releases atoms at a rate $r_d(t)$, and the cell
walls can both release and capture atoms, changing the number of
adsorbed atoms $N_w$:
\be
  p(t)= \alpha(r_d(t) - \dot{N}_w)\,.
\ee
Here, $\alpha$ is a constant that depends on the pumping speed.
$N_w$ is governed by a rate equation,
\be
\dot{N_w}=-N_w\tau_s^{-1}+\beta r_d(t)\,,
\ee
where $\tau_s$ is the mean overall sticking time that an atom spends
on the cell walls before being pumped out, and $0\ge\beta\ge 1$ is the
total probability for an atom to stick to the cell walls
rather than being pumped away directly. In an experiment with constant
alkali release rate $r_d(t)=\bar{r}_d$ (``cw'' experiment), $N_w$ will
reach a steady state $N_s=\tau_s \beta \bar{r}_d$ which balances
desorption and adsorption. Once this steady state has been reached,
pressure no longer depends on the adsorption properties and remains
constant at $p_c=\alpha \bar{r}_d$.  In a pulsed experiment, the cycle
time is typically less than a minute, while $\tau_s$ is expected
to be much longer. Consequently, $N_w$ in the pulsed case will undergo
only small oscillations about a steady-state value $N_s'=\tau_s \beta
\bar{r}_d'$, where $\bar{r}_d'$ is the release rate averaged over one
cycle.  For the purpose of this model, we will neglect the dispenser
time constant and take $r_d'(t)$ to be square pulses, with a constant
high value during the MOT loading phase of duration $T_M'$ and
$r_d'(t)=0$ during the ``science'' phase of duration $T_s'$. The
pressure during the science phase is then $p_s'\approx\alpha
N_s'\tau_s^{-1}=\alpha\beta\bar{r}_d'$.

Now we can compare a cw and a pulsed experiment which both use the
same MOT setup (beam diameters, detunings, field gradients etc.) and
load the same number $N$ of atoms into the MOT. For simplicity we
assume that the alkali pressure dominates over other gases, so that
the required release rate during MOT loading is inversely proportional
to the MOT loading time (see \ref{sec:methods} below for a discussion
of MOT loading). In the cw case, this release rate is kept constant,
so that $\bar{r}_d=\gamma T_M^{-1}$, where $\gamma$ is a constant
depending on $N$ and on the MOT setup. The constant steady-state
pressure is then
\be
  p_c=\alpha\gamma\frac{1}{T_M}   \,.
\ee
For the pulsed case, a rate $\gamma {T_M'}^{-1}$ is applied
during the MOT loading phase, but the rate drops to zero during the science
phase. The average rate is then $\bar{r}_d'=\gamma/(T_M'+T_s')$ and
the pressure during the science phase is
\be
  p_s'=\alpha\gamma\frac{\beta}{T_M'+T_s'}\,.
\ee
The pressure
during the science phase in the pulsed experiment is related to the
constant pressure of the cw experiment by
\begin{equation}
{p_s'} = \beta \frac{T_M}{T_M'+T_s'}p_c\,.
\label{eq:pressureRatio}
\end{equation}
The relation depends on the sticking probability $\beta$. Even for the
worst case $\beta=1$, $p_s'$ is lower than $p_c$ as long as the cycle
time $T_M'+T_s'$ of the pulsed experiment is no shorter than the MOT
loading time of the cw experiment. The ratio $T_M/(T_M'+T_s')$
reflects our assumption $\bar{r}_d\propto 1/T_M$ for a given $N$,
which means that a constant fraction of the released atoms ends up in
the MOT, independently of the rate $\bar{r}_d$ at which the MOT is
loaded. If factors such as non-alkali background pressure change this
relation, the ratio will change. Nevertheless, the qualitative trends
predicted by eq.~\ref{eq:pressureRatio} should still be correct: In a
pulsed experiment, for a given $N$, the pressure in the science phase
increases for higher sticking probability and for shorter cycle
time. Compared to a cw experiment with the same $N$, we expect that
one can always achieve either a faster cycle or a reduced pressure
in the science phase, even in a cell with unfavorable sticking
properties. Achieving both improvements at the same time may also be
possible, but requires the sticking probability to be significantly
below 1.

\begin{figure}
% Fig 1
  \centering
  \includegraphics[width=\columnwidth]{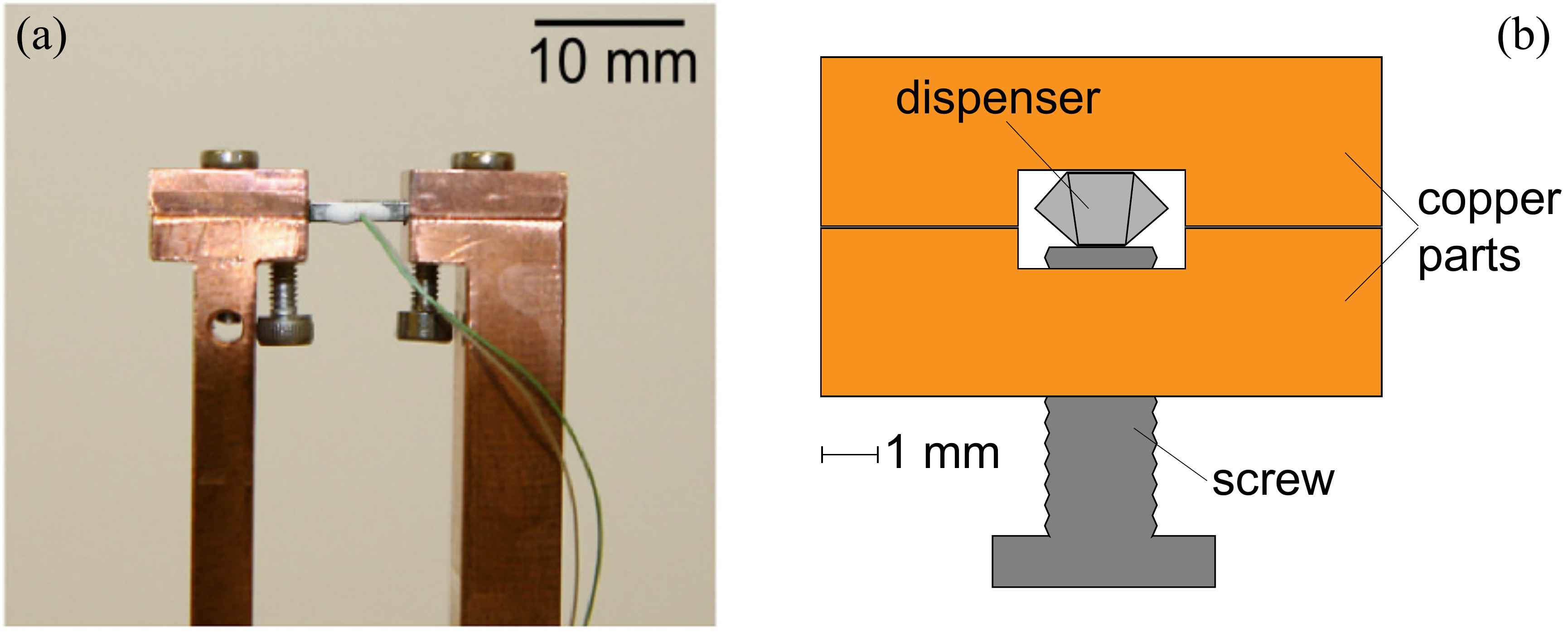}
  \caption{(a) Photo of the dispenser holder. Each of the two copper
    bars provides heat sinking from the dispenser and simultaneously
    serves as an electrical contact. (The thin wires and glue spot on
    the dispenser belong to a thermocouple that was used for thermal
    tests. They were not present on the devices used in the
    experiment.) (b) Schematic cross-sectional view of the copper clamp.}
\label{fig:holder}
\end{figure}

\begin{figure}
% Fig 2
%\includegraphics[height=5cm]{setupPhotoJR2a}
%\includegraphics[height=5cm]{setupPhotoJR2b}
\includegraphics[height=5cm]{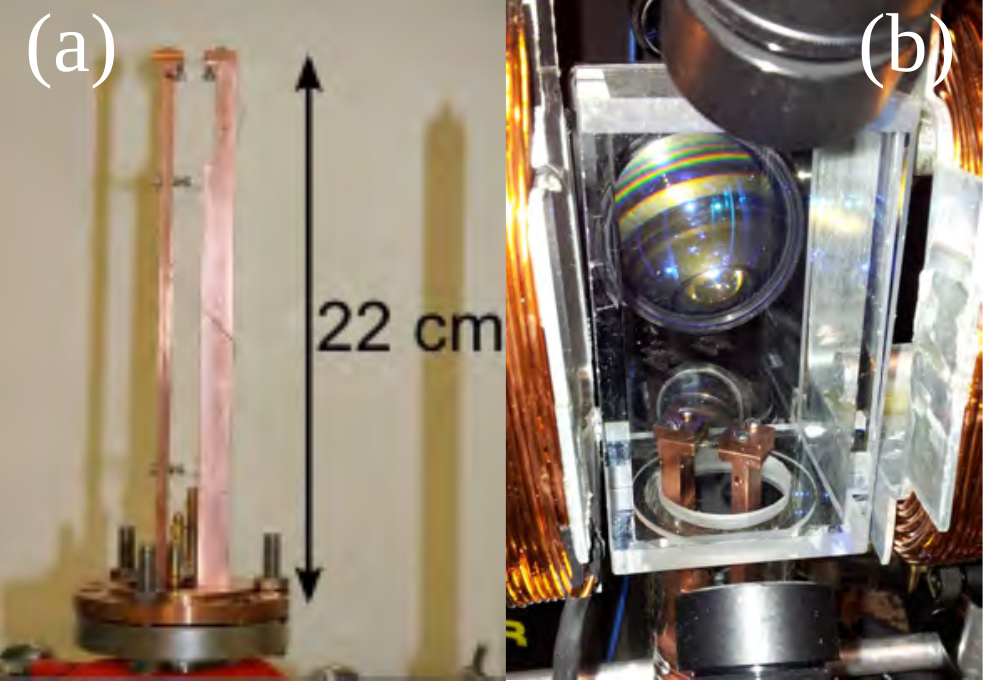}
\caption{(a) Full view of the dispenser holder and copper
  flange. (Again, the thermocouple wires are not part of the final
  version.) The excessive length of the bars was necessary in our
  implementation because of the long glass-to-metal transition of our
  reused glass cell.
% shown in (b).
(b) View of the glass cell containing the dispenser holder. The
fluorescence collection lens is visible behind the cell.}
\label{fig:setup}
\end{figure}

\section{Experiment}
\label{sec:experiment}
\subsection{Setup}
In order to improve the heat flow out of the dispenser, we have
mounted it in copper clamps on the top surface of massive copper bars,
as shown in Figs.~\ref{fig:holder} and \ref{fig:setup}. The clamps
provide a much better thermal contact than the spot-welded or crimped
wires that are usually employed.  The bars (cross section $\sim
1\,\mbox{cm}^2$) provide the thermal connection to a low-temperature
reservoir outside the vacuum system. Ideally, they should be very
short in order to keep their thermal resistance negligible. Our system
is quite far from that ideal (Fig.~\ref{fig:setup}) because we have
reused an existing glass cell which had a particularly long
glass-to-metal transition, so that the rods had to have a total length
of 22\,cm. This reduces the overall thermal conductivity and leads to
a significant temperature rise in the upper part of the rods,
contributing to the long-term behavior described below in
sec.~\ref{sec:longTerm}. In a purpose-built vacuum system, this length
could be made much shorter.  Heat-sinking from the copper bars is
achieved by a copper flange that has the diameter of a standard CF
copper gasket, but a thickness of several centimeters. The copper bars
are screwed to the inside of this flange, while passive heatsinks or
optionally water cooling can be attached to its outer surface.  The
quartz cell of volume 290\,ml is pumped with a 25\,l$/$s ion pump via
a 35CF cross, tube (length 20\,cm), and an angle valve. Taking into
account the tubes and angles, the calculated pumping speed at the cell
entrance is 4\,l$/$s and the pumping time constant about 70\,ms.  The
base pressure as indicated by the ion pump current is below
$10^{-9}$\,hPa.

The MOT is a standard setup with six independent beams. The axial
gradient of the magnetic quadrupole field is 11\,G$/$cm in all
experiments reported here. In our test setup, the $1/e^2$ beam
diameter (1.2\,cm) and the power per beam (1.6\,mW) are both small,
which limits the trapped atom number.  This should be taken into
account when comparing our atom numbers to other experiments: atom
numbers can be expected to increase by several orders of magnitude if
the same dispenser assembly is used with the larger beam diameters and
laser powers that are typical in state-of-the-art quantum gas
experiments.

\subsection{Measurement methods}
\label{sec:methods}
As mentioned above, all the experiments described here were carried out
after several months of operation, i.e. long after the system had
cured.  Fluorescence light from the atoms in the beam intersection was
collected by a 2\,in diameter, $f=40\,$mm lens outside the glass cell,
followed by an amplified photodiode. Unless otherwise indicated, the
fluorescence was measured during MOT loading.  Because of the
uncertainties in solid angle and in the average dipole strength of the
fluorescing atoms, the absolute values of our atom number and loading
rate measurements are beset with a large systematic error.

To deduce the Rb pressure from the fluoresce during MOT loading, we
use the well-established loading rate analysis (see for example
\cite{Monroe90,Mimoun10,Arpornthip12}). In our regime of moderately
low trapped atom number $\Nmot$, MOT loading and decay are well
described by

\be
 \dot{\Nmot}=R-\frac{\Nmot}{\tauMot}\,.
\label{eq:MotEq}
\ee 

$R=\alpha_R\pRb$ is the MOT loading rate, $\pRb$ the partial pressure
of \Rb\ and $\alpha_R$ is a constant that depends on the parameters of
the MOT setup (beam diameter, detuning and intensity and magnetic
field gradient).  $\tauMot$ is the decay time constant due to
collisions with the background gas. It is proportional to the sum of
the partial pressures of \Rb\ and of all contaminants $i$ present in
the vacuum chamber, each weighted with the relevant collisional cross
section $\sigma_\text{Rb-i}$.  We separate loss due to thermal Rb
atoms from the loss that is independent of Rb pressure:

\be
 \tauMot^{-1}=\betaRb\pRb+\tau_b^{-1}\,,
\ee

When pressures are constant, eq.~\ref{eq:MotEq} leads to
\be
  \Nmot(t)=\Nfull(1-e^{-\frac{t}{\tauMot}})
\label{eq:MotLoad}
\ee

with

\be
\Nfull=R\tauMot=\frac{\alpha_R \pRb}{\betaRb\pRb+\tau_b^{-1}}\,.
\ee

\Rb\ pressure is proportional to $R$. The background pressure is
roughly proportional to $\tauMot^{-1}$ because the collisional
cross-sections of background gas constitutents are all comparable
\cite{Arpornthip12}. Thus, both quantities can be deduced from
$\Nmot(t)$. During the dispenser pulse, where pressures vary too fast
for this solution to be valid, $R$ can still be measured from the
initial linear rise in $\Nmot(t)$.

% Because we are interested in UHV
% systems where the base pressure corresponds to background collision
% rates well below 1\,s$^{-1}$, we assume $\Ppart$ to dominate during
% MOT loading. Thus, two quantities of interest can be deduced from a
% measured loading curve: the initial MOT loading rate is proportional
% to the \Rb\ pressure and the time constant $\tau$ is the
% background-limited lifetime of ``typical'' trap.

\section{Results: Pressure modulation}
\label{sec:shortTerm}

\begin{figure}
% Fig 3
\includegraphics[width=\columnwidth]{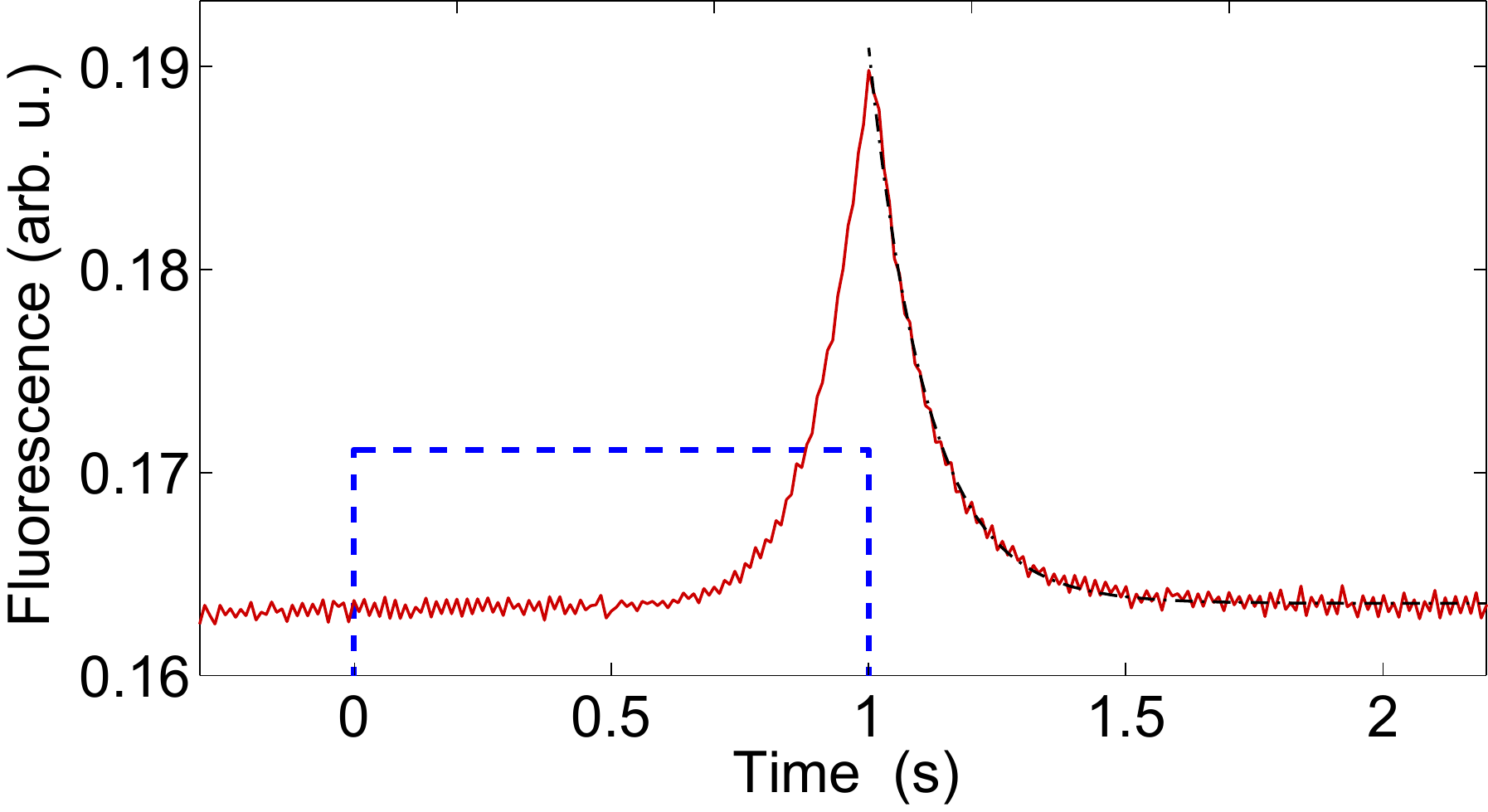}
\caption{Pulsed dispenser emission observed in fluorescence, MOT coils
  switched off. A
  current of $I_d=20.2\,$A is applied for 1\,s starting at $t=0$ (blue
  dashed curve). The red solid curve shows the fluorescence observed
  (signal averaged over 29 such pulses). This data was taken in steady state
  (i.e., after many identical pulse cycles).  The decay time constant
  of the exponential fit (black dot-dashed line) is $\tau_1=112\,$ms.
\label{fig:singleFluo}}
\end{figure}

\begin{figure}
% Fig 4
\includegraphics[width=0.7\columnwidth]{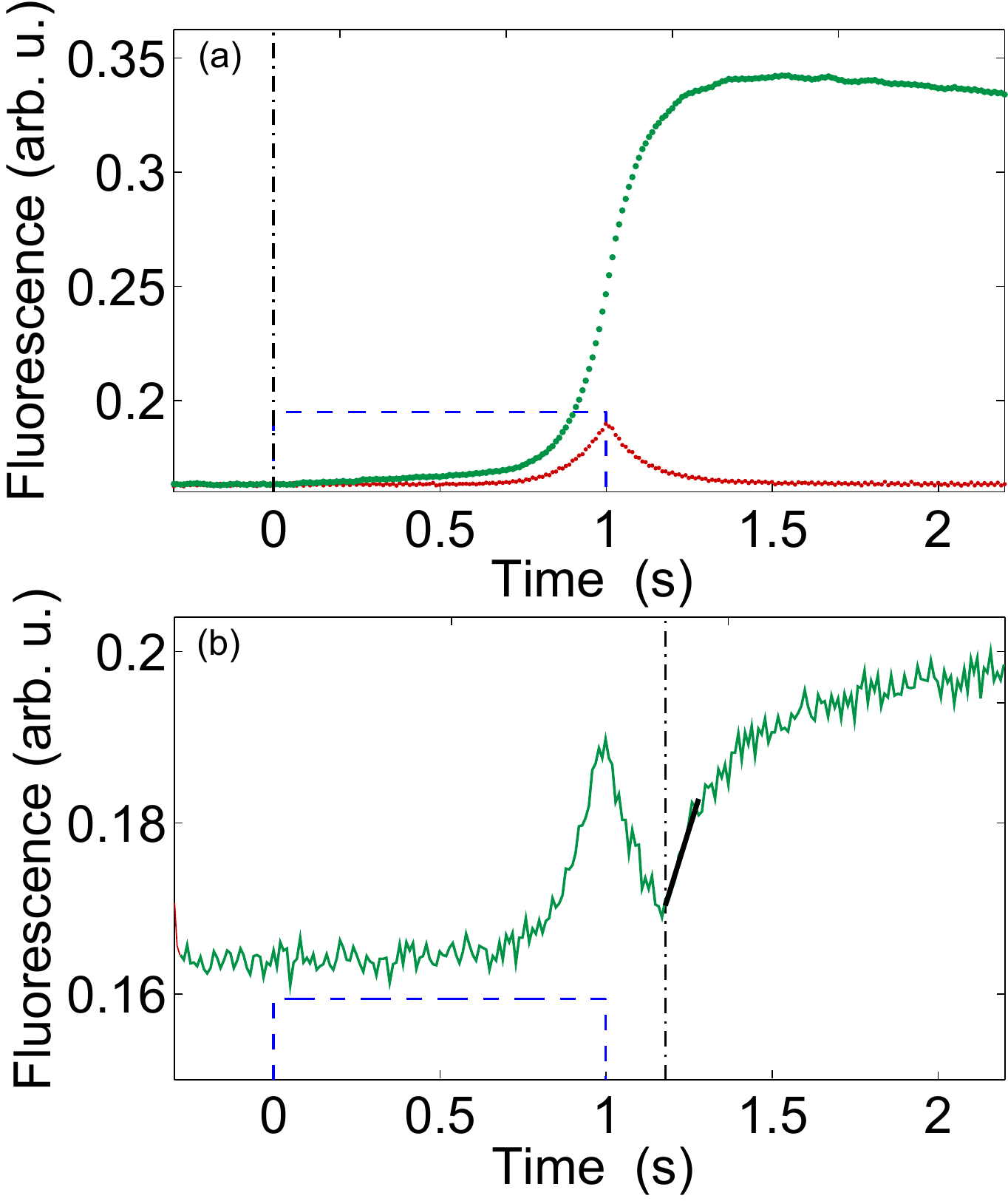}
\caption{MOT fluorescence (green curves) during loading from the
  pulsed dispenser. Blue dashed curves indicate the timing of the
  current pulse (start at $t=0$, end at $t_\text{end}= 1\,$s,
  $I_d=20.2\,$A).  Black dashed lines show the start of the MOT
  loading (switching on the MOT coils), which occurs at $t=0$ in (a)
  and at $t=t_\text{end}+0.19\,$s in (b). The MOT signal in (a) is
  averaged over 10 shots, the signal in (b) is a single shot. The red
  curve in (a) recalls the time dependence of the rubidium vapor
  density in the cell (same data as in Fig.~\ref{fig:singleFluo}). The
  black solid line in (b) shows the initial slope of the MOT loading
  curve, which is used to deduce the pressure of \Rb\ in the cell at
  that time.}
  \label{fig:singleMot}
\end{figure}

\begin{figure}
% Fig 5, was 6 but Peter wanted it to appear earlier
\includegraphics[width=0.7\columnwidth]{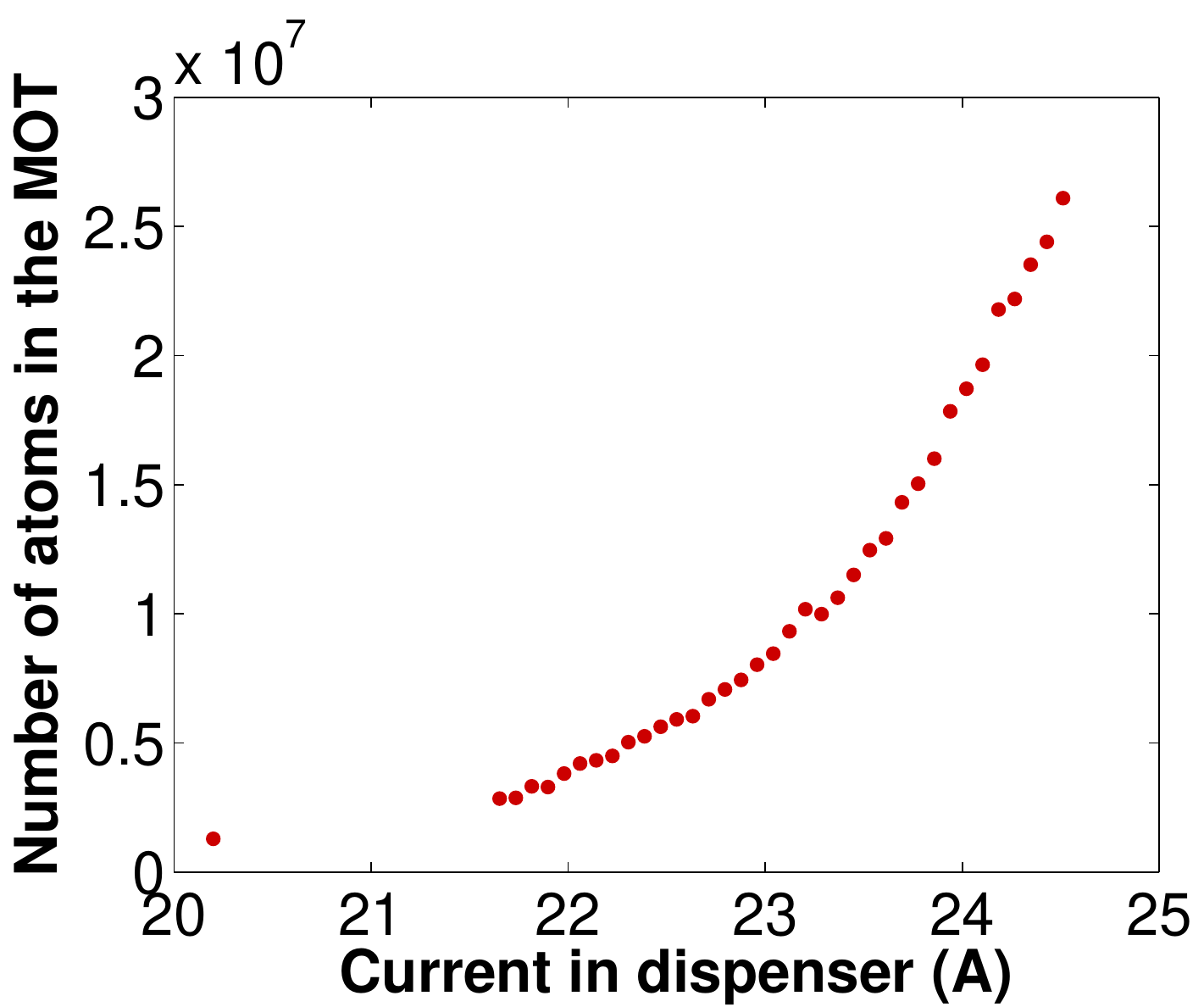}
\caption{Number of atoms in the MOT as a function of dispenser current
  $I_d$. Current pulses had a duration of 1\,s. MOT coils were
  switched on at the start of the current pulse and the atom number
  was measured 200\,ms after the end of the pulse.}
\label{fig:NvsId}
\end{figure}

\begin{figure}
% Fig 6, was 5
\includegraphics[width=0.95\columnwidth]{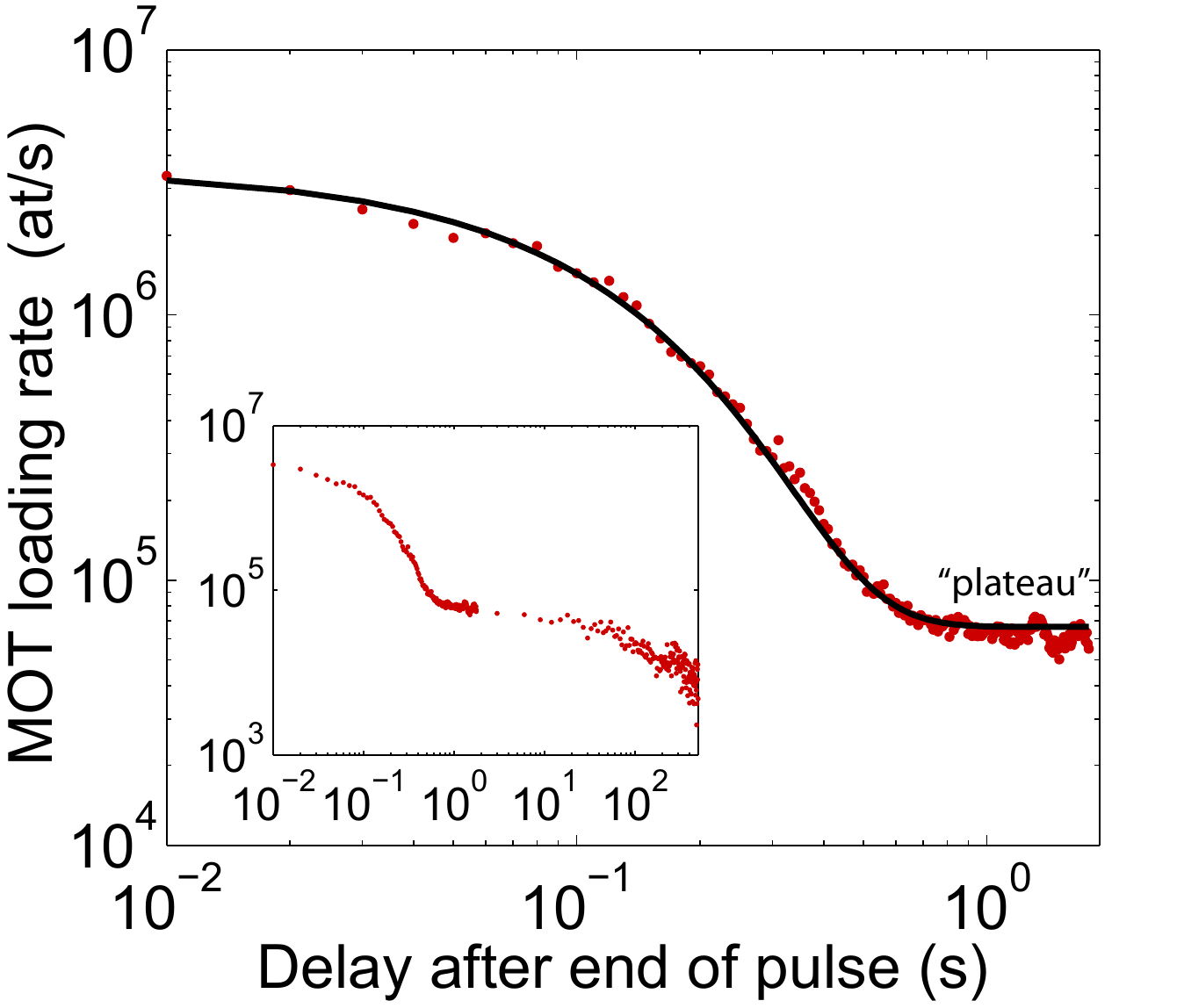}
\caption{\Rb\ pressure as a function of the time $t_0$ after the end
  of a $1\,s$ dispenser pulse with $I_d=20.2\,$A. The initial loading
  rate of a MOT started at $t_0$ is used to measure the pressure.  To be
  sure that steady-state conditions were reached, these measurements 
  was taken after 10 hours of pulsed operation. The
  black solid line in the main graph is an exponential fit with offset as
  described in the text, which
  yields a decay time constant of 108\,ms, in good a agreement with
  the fluorescence measurement above. Rb pressure decays by a factor
  of 51 during this fast decay. The inset shows data on a longer
  timescale, revealing a second, slower decay on the scale of
  100\,s. 
  \label{fig:loadingRate}}
\end{figure}

In a first experiment, we measure \Rb\ fluorescence emanating from the
beam intersection while applying short current pulses to the
dispenser. The MOT magnetic field remains off.
% Due to residual magnetic fields and beam imbalance,
Therefore, the photodiode signal is proportional to the instantaneous
pressure of room-temperature \Rb\ in the cell (with some integration
due to a short-lifetime optical molasses created by the
beams). Applying 1\,s pulses with various currents $I_d$, we find that
the threshold for \Rb\ emission lies around
$I_d=18\,$A. Fig.~\ref{fig:singleFluo} shows the fluorescence for
$I_d=20.2\,$A. The rubidium pressure decay after the end of the pulse
is well described by an exponential decay, yielding a time constant of
$\tau_f=112\,$ms. This decay time is at least 20 times shorter than
the values reported for bare dispensers \cite{Rapol01,Bartalini05},
confirming the validity of our approach.  We have checked that the
time constant depends only weakly on the value of $I_d$ and on the
system temperature (see section~\ref{sec:longTerm} below).  We also
observe that very little \Rb\ is emitted during the first $0.7\,s$:
the dispenser heats up, but has not yet reached emission
temperature. After this delay, the pressure rises very quickly to very
high values. It would be possible to further accelerate the emission
by using a higher current until emission starts and then lowering it,
instead of the simple square pulses we use here.

% The parameters of the figure correspond
% to the experimentally most suitable range for our dispenser mount: the
% current is high enough to reach full MOT loading in less than a
% second; yet, the combination of current and pulse duration remains in
% a safe region with sufficient headroom to avoid dispenser destruction.

Next, we have used these rubidium pulses to load the MOT. In
Fig.~\ref{fig:singleMot}(a), we use again $I_d=20.2\,$A and 1\,s pulse
duration, and the MOT coils are turned on together with the current
pulse at $t=0$. Full loading takes about 1.2\,s, including the 0.7\,s heating
time where no \Rb\ is emitted. As can be seen in the figure, most
of the loading takes place during the final 0.5\,s. It may be possible to
further accelerate the cycle by starting the heating before the end of
the preceding cycle. (We have not attempted to do this here.)

We have measured the number of trapped atoms as a function of
dispenser current $I_d$. As before, current pulses were 1\,s long and
the MOT coils were turned on at the start of each pulse. 
Because the photodiode also receives
fluorescence from untrapped background \Rb\ atoms (visible as a peak
before MOT switch-on visible in fig.~\ref{fig:singleMot}b), 
$\Nmot$
was measured 200\,ms after the end of the pulse. (This time is long
enough for background fluorescence to become small, while being
shorter than MOT lifetimes at these pressures.)
% \jrcommt{Incorporate info from Vincents email of 27.10.13: in what
%   sense is this ``fully loaded''...} As we increase $I_d$, the number of atoms $\Nfull$ in the fully-loaded
% MOT increases and time to full loading becomes shorter, shifting
% towards the end of the current pulse \jrcommt{Correct, Vincent? You
%   said that you observed this acceleration in the numerical
%   simulation, but do you also see it in the
%   experiment?}.
$\Nmot$ increases with $I_d$ and shows no saturation up to the highest
currents that we dared to apply (Fig.~\ref{fig:NvsId}). This suggests
that the pressure of elements other than rubidium remains
non-negligible even at the highest currents, but rises more slowly
than $\pRb$. 
% \jrcommt{This could be due to outgassing. Vincent, do you
%   remember in which order you took the data? This might give an
%   indication: if you started with the low currents, there would be an
%   effect of outgassing at higher and higher temperatures, so the dirt
%   pressure would be expected to rise as $I_d$ increases. But if you
%   started with the high currents, most of the outgassing should have
%   occured in the beginning, and this explanation is less likely.}
% Pour ta question sur l'ordre dans lequel les donnees ont ete prises:
% je les ai prises par ordre de courant croissant. cependant ici le
% degazage ne joue qu'un role mineur, puisque je n'ai produits que tres
% peu de pulses (qql dizaines), en un temps assez long (ces donnes ont
% ete prises au debut, lorsque je n'avais pas de control par carte NI,
% donc j'ai du modifier les reglages a la main pour chaque courant
The absolute numbers of $\Nmot\sim 2.5\cdot 10^7$ for the highest
currents compare favorably to those of steady-state MOTs with similar
beam diameters such as the TACC experiment in our lab \cite{Deutsch10}.

Measuring the initial linear slope of MOT loading curves provides an
alternative and more sensitive way to measure $\pRb$. We use a
dispenser pulse duration of 1\,s as before. We record fluorescence
curves for various delays $t_0$ between the end of the current pulse
and the start of MOT loading, which is initiated by switching on the
coils. After each loading cycle, MOT coils are switched off for 0.3\,s
to purge all cold atoms.  The measurements are analyzed by performing
a linear fit to the fluorescence curve in a 100\,ms time interval at
the start of the MOT loading. Its slope yields the loading rate $R$ at
$t_0$. To avoid measurement errors due to the background fluorescence
mentioned above, we limit our analysis to times after the dispenser
switch-off, where we have checked that the variation of background
fluoresce is only a small contribution to the measured slope. (To
obtain loading rates for earlier times, it would be possible to
subtract the time-dependent background fluorescence, which can be
measured independently by simply leaving the MOT coils off.)
Fig.~\ref{fig:singleMot}(b) shows an example fluorescence curve during
one such cycle for $t_0=0.19\,$s.  The curve $R(t_0)$ resulting from
many such measurements is shown in Fig.~\ref{fig:loadingRate} on a
log-log scale. To be sure that a steady state was reached, this
measurement was taken after 10 hours of pulsed operation. As expected,
the curve shows an evolution with two well-separated time
scales. Pressure drops exponentially by a factor of 50 with a fast
time constant of $\tau_1=108\,$ms. This time constant is in good
agreement with the value obtained by the simple fluorescence
measurement above.  It then reaches a plateau, from which it decreases
by at least another order of magnitude on a much slower time
scale. Doubly exponential decay was observed in LIAD experiments
\cite{Anderson01,Klempt06}. In our case, the slow decay is not well
approximated by an exponential. Its timescale is roughly 200\,s. We
attribute it to a combination of cell heating and
adsorption-desorption dynamics. (If the wall temperature was constant,
it would allow us to estimate the sticking time $\tau_s$ of
sec.~\ref{sec:adsorption}.)  The long-term pressure
evolution is studied in more detail in the next section.

%Hier war "We have measured the nb of trapped atoms as fct of I"

To summarize these results, the rapid pressure drop by a factor of 50
in about $100\,$ms demonstrates the usefulness of this source for
high-repetition-rate experiments. It confirms that dispenser cooldown
was the limiting process for the slow pressure decay in previous
pulsed-dispenser experiments, which used bare dispensers. In other
words, in those experiments, source shut-off was much slower than the
pumping processes. Although the pressure plateau reached after the
fast decay is above the base pressure in the cell, the pressure
modulation provides a significant advantage over the unmodulated
dispenser. This becomes clear when comparing to a single-cell
experiment with unmodulated dispenser, such as TACC
\cite{Deutsch10}. TACC has somewhat larger MOT beam diameters
(1.8\,cm), a similar base pressure, and its constant Rb pressure is
similar or higher than the pressure plateau observed here (loading
rate $R\approx 3\times 10^6$at$/$s with $\tauMot=4.8\,$s).  Yet, full
MOT loading takes more than 10\,s in TACC instead of 1.2\,s here, and
achieves an atom number that is at best comparable, and typically
lower than observed here.

In the next section, we will try to get a better understanding of the
plateau, and of the long-term evolution of the pressures in general.

\section{Results: Long-term behavior}
\label{sec:longTerm}

\begin{figure}
% Fig 7
\includegraphics[width=0.8\columnwidth]{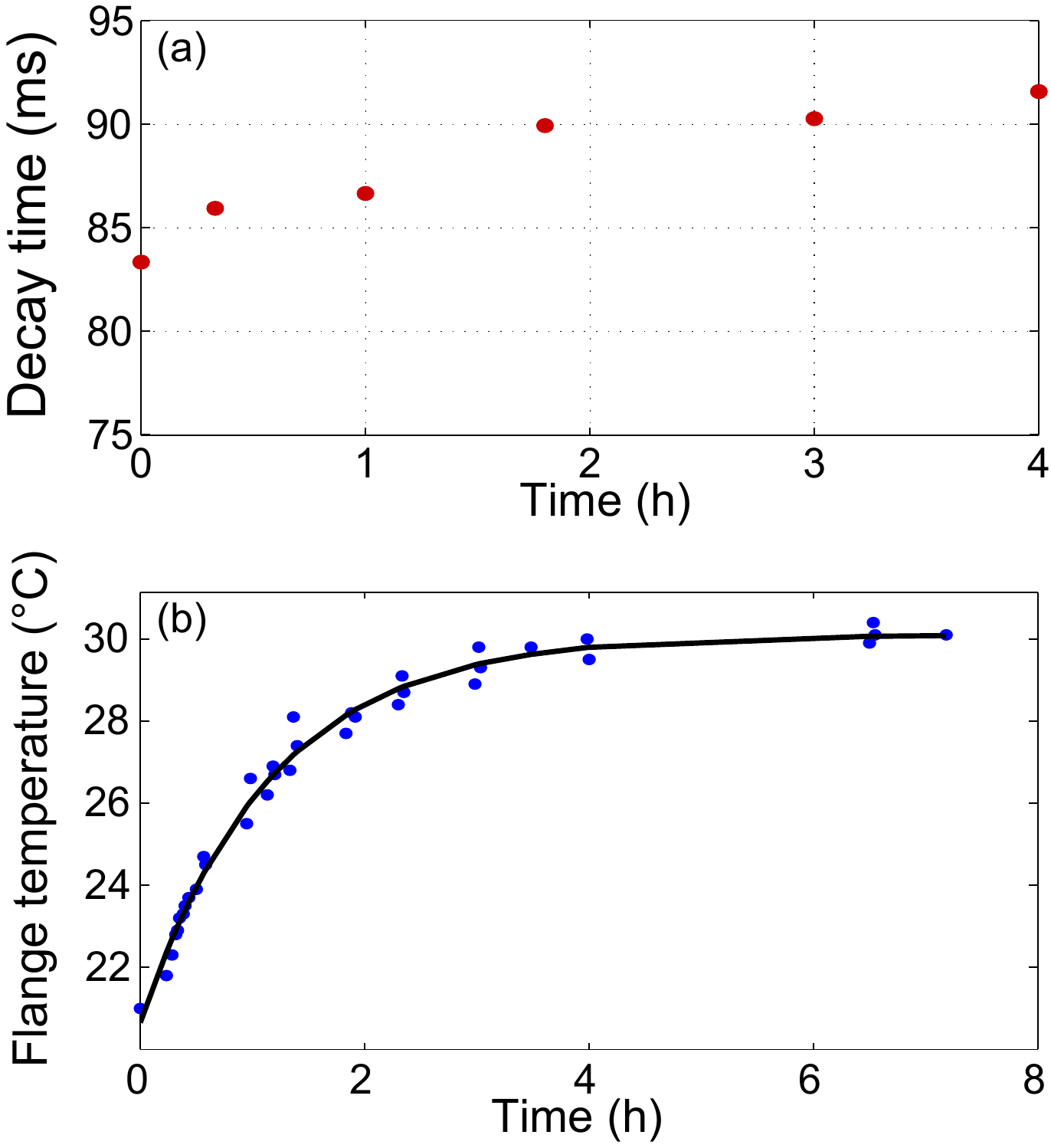}
\caption{(a) Long-term evolution of the pressure decay time constant
  $\tau_f$. Current pulse with $I_d=21.2\,$A and 1\,s duration were
  applied every 5\,s. The decay time drifts by about 10\% and settles
  after about two hours. (b) Temperature of the copper flange during a
  similar experiment. The time constant of the exponential fit (solid
  line) is 1.2\,h, similar to the settling time in (a).}
\label{fig:longTermFluo}
\end{figure}

\begin{figure*}
% Fig 8
\includegraphics[width=0.8\textwidth]{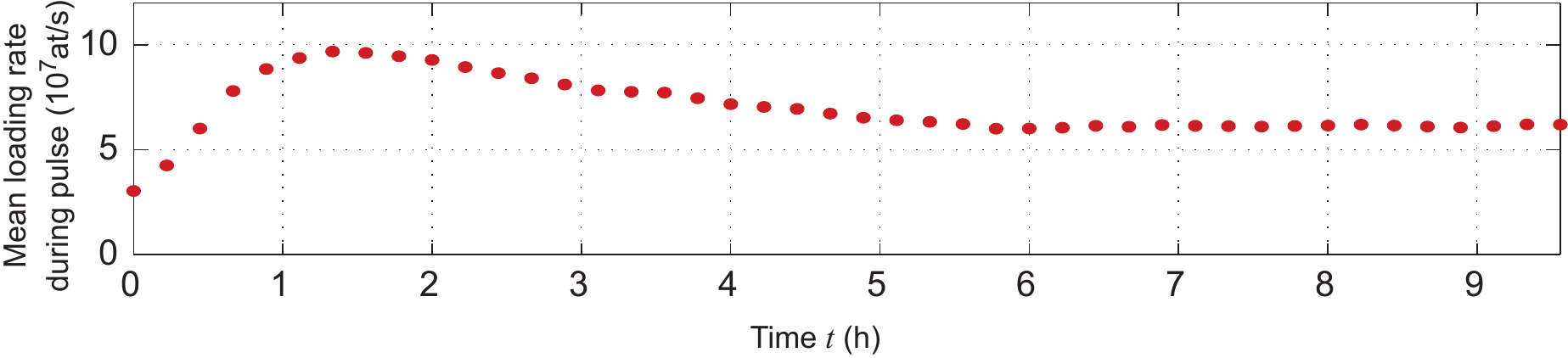}
\caption{Long-term evolution of the mean MOT loading rate during the
  dispenser pulse. For $t<0$, the dispenser was permanently off for
  several hours (``cold start'' at $t=0$).  Starting from
  $t=0$, 1\,s pulses with $I_d=24\,$A were applied every 5\,s. 
% Ich hatte vorher I=22.8A aus irgendwas geschlossen. Vincent
% berechnet 24A aus der Kontrollspannung, die er im Laborbuch gefunden hat.
}
\label{fig:loadingRateLongTerm}
\end{figure*}

First, we examine the long-time stability of the pressure decay time
constant $\tau_f$.  As the Rb release from the dispenser is a very
nonlinear function of temperature, $\tau_f$ might have a strong
dependence on the temperature of the copper bars, which would be
problematic because this temperature is difficult to control
precisely. In this respect, our long copper bars represent a
worst-case scenario: due to their length which limits heat evacuation,
the temperature of the copper structure rises slowly
during operation, and with it the mean dispenser temperature will rise
as well. Therefore, we have repeated the decay time measurement of
Fig.~\ref{fig:singleFluo} every 5\,s over a duration of 4 hours, using
a current $I_d=21.2\,$A. Fig.~\ref{fig:longTermFluo}(a)
shows the result: the time constant undergoes an increase that is
measurable, but remains below 10\%. It occurs during the first 2\,h;
for longer times, $\tau_f$ remains constant within the error margin. We
have also measured the evolution of the temperature at the base of the
copper mount during a similar measurement
(Fig.~\ref{fig:longTermFluo}(b)). The temperature rises by about
$9^\circ$C and settles with a time constant of about 1.2\,h,
comparable to the settling time of $\tau_f$. This suggests that the
temperature change might indeed be responsible for the change in
$\tau_f$.  The steady-state value of 91\,ms observed in this
measurement is about 20\% shorter than the one observed in
Fig.~\ref{fig:singleFluo} for $I_d=20.2\,$A. None of these variations
will cause major perturbations for fast trap loading.

Another quantity of interest is the long-term evolution (over many
pulses) of the peak \Rb\ pressure reached during the pulse. This determines
the long-time evolution of the MOT loading rate and of the total
number of trapped atoms per cycle. To measure it, we again apply a
1\,s current pulse every 5\,s over a long time (several hours to
days).
%Dispenser current is $I_d=?$. \jrcommt{What is
%  the current value?} 
The MOT is switched on at the start of each
pulse, and switched off for 300\,ms before the next one to
purge all cold atoms. We record the MOT fluorescence and deduce a mean
loading rate during each pulse by dividing the total number of atoms
%\jrcommt{at what time precisely? Or is it the peak value?} 
measured 200\,ms after the end of the current pulse by the useful
pulse duration of 0.2\,s (see Fig.~\ref{fig:singleFluo}). The result
is shown in Fig.~\ref{fig:loadingRateLongTerm}. The loading rate rises
during the first 1.5\,h. to a peak value which is about 1.7 times
higher than the steady-state value. The steady state is reached after
5 to 6\,h, which is rather long.  
%besser formmulieren: 
This should not be a problem in practice: long time constants for
pressure stabilization also appear when turning on a cw
dispenser. Just as one keeps the cw dispenser switched on the whole
day, the dispenser pulsing should be maintained the full day, even
when the experiment is not being used for a while. Nevertheless, we
would like to better understand this phenomenon, which is the subject
of the next paragraphs.

\begin{figure}
% Fig 9
\includegraphics[width=0.8\columnwidth]{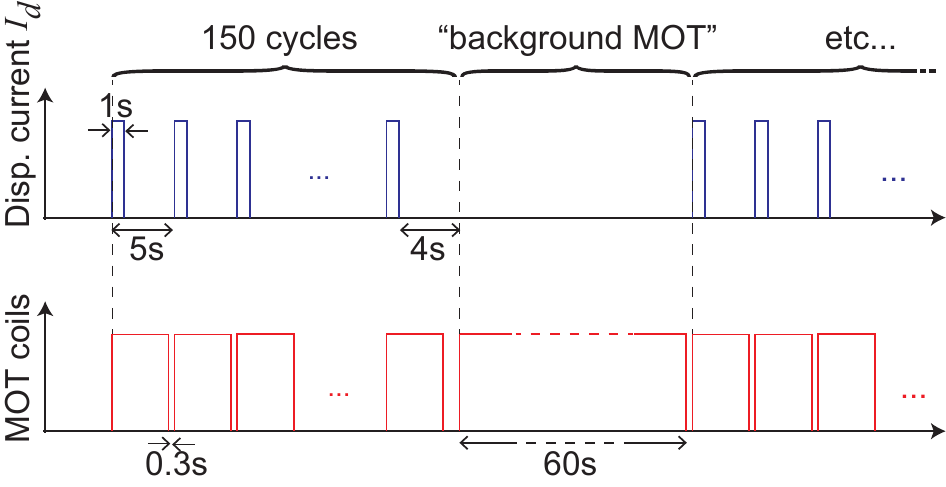}
\caption{Timing sequence for measuring long-term evolution.}
\label{fig:timing}
\end{figure}

\begin{figure*}
% Fig 10
\includegraphics[width=0.8\textwidth]{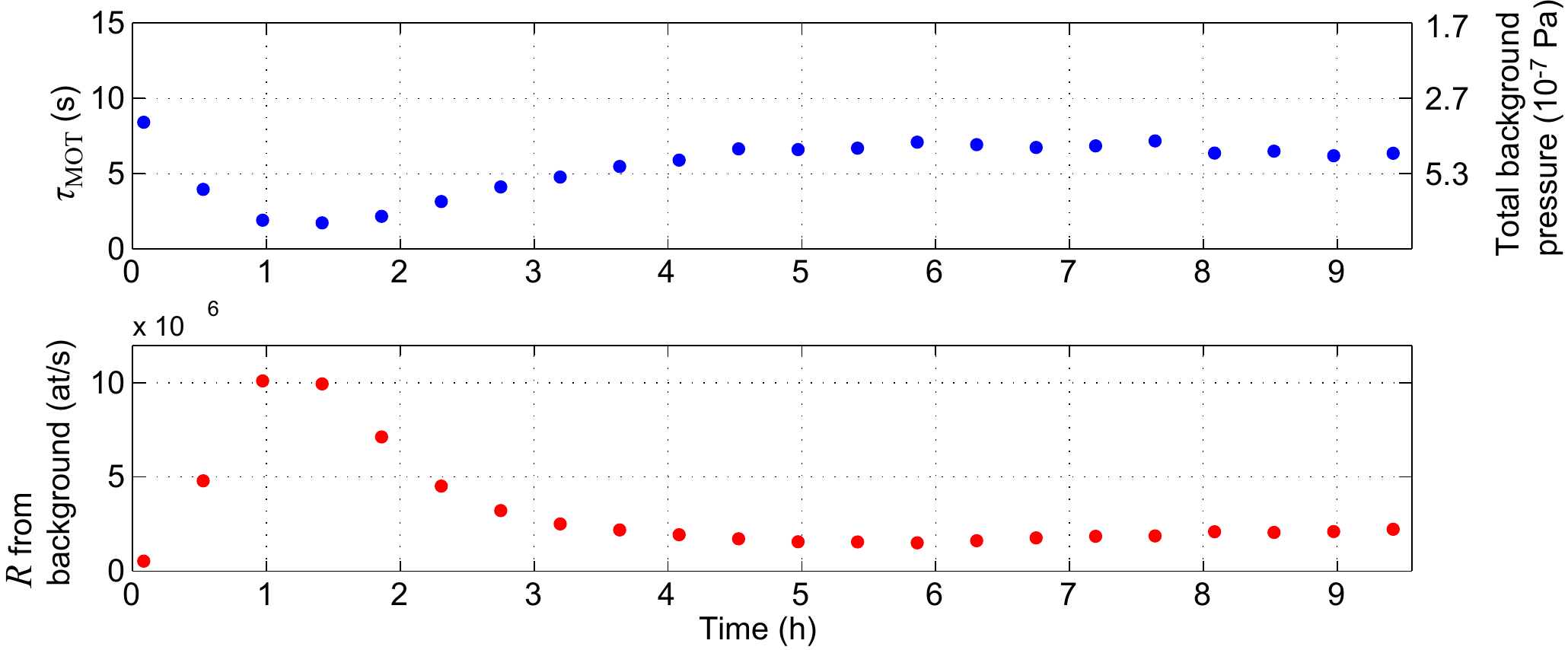}
\caption{Long-term evolution of the low-pressure plateau. This data is
  obtained using the ``Background MOT''
  (cf.~Fig.~\ref{fig:timing}). $\tauMot$ is a measure of the total
  pressure, while $R$ is proportional to the partial pressure of
  \Rb. The plateau pressures follow a pattern similar to the pulsed
  loading rate.}
\label{fig:pressuresLong}
\end{figure*}

Two mechanisms introduce long time constants in our system. The first
is caused by the slow heating due to excessive length of the copper
bars, discussed and characterized above. The heating time constant is
compatible with the pressure peaking time in
Fig.~\ref{fig:loadingRateLongTerm}. A mechanism that could cause the
loading rate to peak and then decay in this scenario is release of
previously adsorbed Rb from the copper surfaces. This effect will be
substantially reduced in an improved mount with shorter copper bars.

The second, more fundamental effect is the adsorption-desorption
dynamics of Rb on all internal surfaces of the cell. On the typical,
cured surfaces, binding energies of Rb are much lower than on
freshly cleaned ones. Hence, Rb atoms adsorbed to such a surface do not
remain there forever, but are released after a time that can be on
the scale of hours or days. Indeed, it is an empirical
fact that Rb MOTs can still be loaded ``from the background'' hours
and days after a cw dispenser has been switched off
\cite{Wieman95}. Moreover, the sticking probability and sticking time
depend on the amount of Rb coverage already present on the
surface. 

Although it is not clear how the adsorption-desorption dynamics would
explain the observed peak in the loading rate, these dynamics
certainly play a role in the long-term evolution of the low-pressure
plateau that is reached a few hundred ms after each dispenser pulse,
and which will determine the lifetime of a magnetic or dipole trap
with our source. To measure the plateau pressure, we insert measurements of
MOT loading without dispenser pulse into our experiment cycle as shown in
Fig.~\ref{fig:timing}: after 150 of the normal 5\,s cycles, we
interrupt the repetitions, wait 4\,s 
% 4s: info Vincent 17.12.12
for the pressure to reach its plateau, and then record a MOT loading
curve for a long time (60\,s) with the source switched off
(``background MOT''). After this, the next group of 150 normal cycles
starts. The purpose of the background MOT is to obtain the plateau
values of \Rb\ pressure and of the total pressure, and their long-term
evolution. The loading curve of the background MOT is fitted with
eq.~\ref{eq:MotLoad} to extract the loading rate $R$ and the loading
time $\tauMot$. $R$ is proportional to the \Rb\ partial pressure in
the plateau, while $\tauMot$ can be used to estimate the total
pressure using the conversion constant $2.7\times 10^{-6}\,\mbox{Pa
  s}$ established in \cite{Arpornthip12}.

Fig.~\ref{fig:pressuresLong} shows the long-term evolution of both
quantities, measured over 9.5\,h.  The curves reveal that the
pressures follow a pattern similar to the pulsed loading rate, with an
extremum after 1\ldots 1.5\,h and a steady state achieved after about
6\,h. The plateau pressure of \Rb\ reached in the extremum is about a
factor of 5 higher than the value reached in the steady-state. Total
pressure and \Rb\ pressure show the same trends. This is not
sufficient to establish the composition of the background gas, but at
least suggests that the variations have a common origin. Once again,
we expect this effect to be mitigated by using shorter copper rods.
%\jrcommt{Should we add the comparison with/without flange cooling?}

\section{Systematic study as a function of the loading rate}
\label{sec:loadingRate}

\begin{figure}
% Fig 11
\includegraphics[width=0.7\columnwidth]{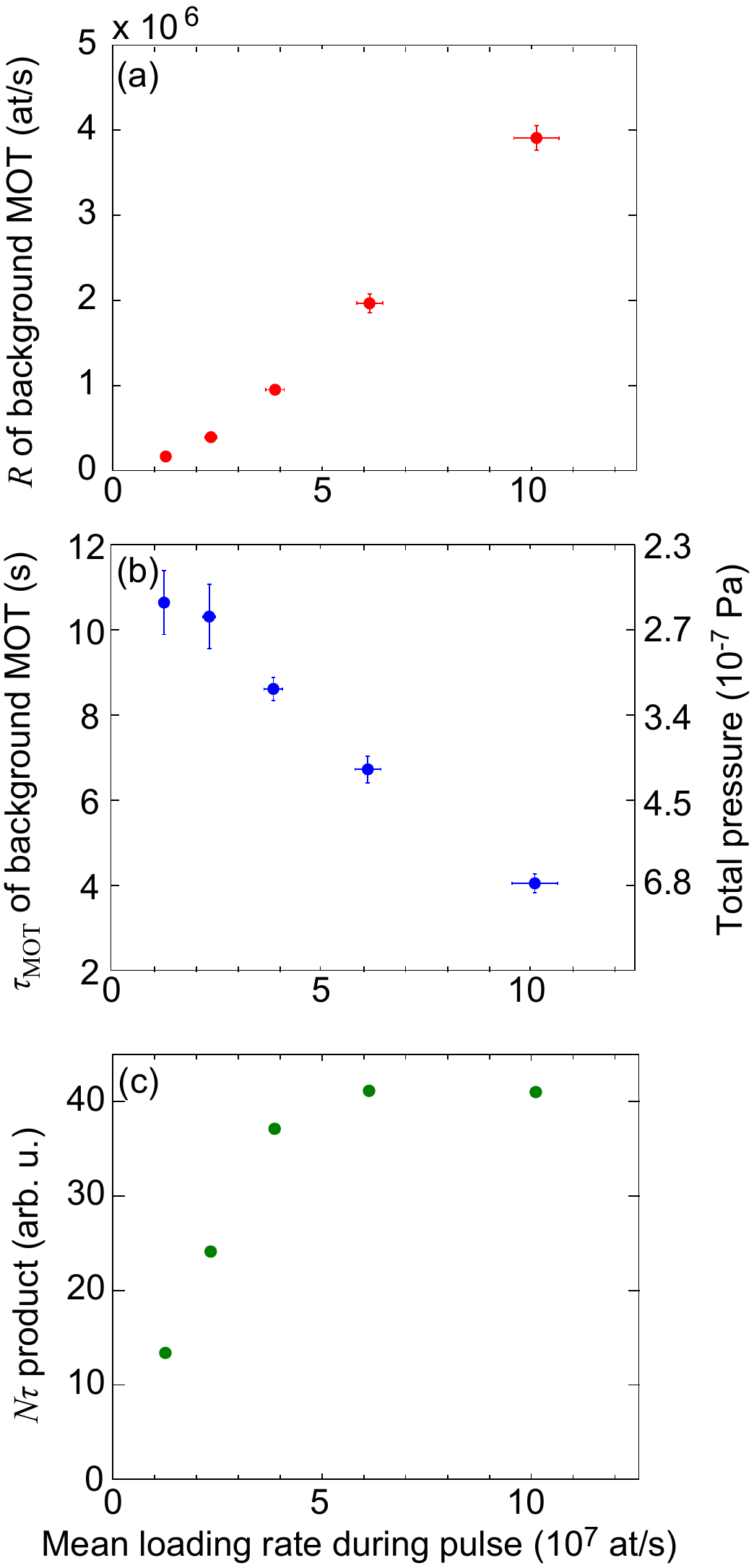}
\caption{The currents of the pulses are $I_d=22.8\,\mbox{A},23.4\,\mbox{A},24.0\,\mbox{A},24.7\,\mbox{A},25.3\,\mbox{A}$.}
\label{fig:summaryModul}
\end{figure}

%The results in the preceding section show that 
%As discussed in \ref{sec:adsorption} and obser the previous section, 
When the transient effects observed in the preceding section have
ceased, the plateau pressure reaches a steady-state value, and this is
the value that is most important to applications. Because of the
mechanisms generating the plateau -- wall adsorption and heating --
the steady-state plateau pressure depends on the amount of \Rb\
released per pulse, and thus on the dispenser current $I_d$.  Using
again the measurement pattern of Fig.~\ref{fig:timing}, we have
measured the total and \Rb\ plateau pressures, as well as the loading
rate during the pulse, for different values of $I_d$. For each $I_d$,
measurements were performed over many hours to obtain the steady-state
value. The results are shown in Fig.~\ref{fig:summaryModul}. As
expected, both the loading rate during the pulse and the plateau
pressures increase with the pulse intensity.

  The figure also shows the product of
atom number and lifetime, which is a figure of merit for evaporative
cooling \cite{Anderson94}. This figure grows for loading rates up to
$5\times 10^{7}$at$/$s and saturates for higher rates.

Let us use Fig.~\ref{fig:summaryModul} to compare again to a cw
experiment. As above, we will compare to the TACC experiment
($R\approx 3\times 10^6$at$/$s, with $\tauMot=4.8\,$s and 
$N=1.4\times 10^7$, achieved after 10\,s of loading). With the
pulsed dispenser, using an intermediate loading rate of $5\times
10^{7}$at$/$s, the atom number is increased by more than a factor 2
and the lifetime by 1.5, while reducing the loading time by a factor
of 10.
%Because
%it has similar MOT beams and base pressure, , is similar to the $3\times 10^7$

%\jrcommt{Is water cooling used in the data of the figures?}
% Vincent: Yes
%\jrcommt{To establish the link to the short-term section: Compare the
%  steady-state values observed here to the values of the
%  short-term section.}

\section{Conclusion}

The pulsed dispenser device demonstrated here provides alkali pressure
modulation by more than an order of magnitude on the 100\,ms timescale
in both directions. While fast pressure rise with pulsed dispensers
has been demonstrated before, the important new feature of our device
is fast pressure decay after the pulse. Although the plateau
pressure reached after this decay is above the base pressure, it is
lower than the cw pressure in a similarly-sized
nonpulsed MOT, while loading is at least an order of magnitude
faster. This performance is achieved in a single-cell setup with only
marginal added complexity over a standard MOT setup. These features
make our device attractive for many applications of ultracold atoms,
including single-cell Bose-Einstein condensation experiments with
sub-second repetition rate and cold atom sensors for field use.

Several further improvements are possible. Our test setup had small
beam $1/e^2$ beam diameters of 1.2\,cm.  Increasing this beam size
would dramatically increase the number of trapped atoms
\cite{Gibble92} for the same alkali pressure, without any changes to
the dispenser device itself. Using current pulses with an initial
``boost'' instead of square pulses would reduce the delay before
alkali emission, further reducing the loading time. Total loading times
below 300\,ms are realistic with this improvement alone
(cf.~Fig.~\ref{fig:singleFluo}).  As discussed in
sec.~\ref{sec:experiment}A, shortening the copper holding bars should
significantly reduce their temperature rise and the associated
transient behavior. Furthermore, having shown that adsorption dynamics
is an important factor for the plateau pressure, we expect that this
pressure can be further reduced by improving the cell geometry and
possibly by a suitable choice of cell materials. Our pulsed dispenser
can also be combined with a nozzle such as described in
\cite{McDowall12} to direct atoms more efficiently towards the capture
region. This may reduce the overall amount of alkali vapor released in
every pulse. Finally, the pulsed dispenser can be combined with
light-induced desorption (LIAD) to further increase the peak pressure
and reduce the surface coverage during the low-pressure intervals.  In
a first test, we have observed a reduction of the \Rb\ pressure in the
science phase by roughly $1/3$ when applying LIAD pulses
simultaneously with dispenser pulses.

\begin{acknowledgments}
  We thank Fabrice Gerbier
  for fruitful discussions. This work was supported in part by the
  Agence Nationale pour la Recherche under the ``CATS'' project.
\end{acknowledgments}

% If in two-column mode, this environment will change to single-column format so that long equations can be displayed. 
% Use only when necessary.
%\begin{widetext}
%$$\mbox{put long equation here}$$
%\end{widetext}

% \begin{figure}
% \includegraphics{}%
% \caption{\label{}}%
% \end{figure}

% \begin{table}
% \caption{\label{} }
% \begin{tabular}{}
% \end{tabular}
% \end{table}

% Create the reference section using BibTeX:
%\bibliography{../../bib/lascool,ArticleDispenserSpecific}
\bibliography{lascool,ArticleDispenserSpecific}

\end{document}